\documentstyle[12pt,psfig]{article}

\newcommand{\C}{C}
\newcommand{\pp}{\partial }

\newcommand{\spc}{\mbox{\hspace{1cm}}}

\def\Tr{{\rm Tr}}
\def\A{{^{(+)}A}}
\def\AA{{^{(-)}A}}

\def\pp{{\partial}}
\def\ket#1{\left\vert #1\right\rangle}

\begin{document}

{\flushright{\small Imperial/TP/97-98/23\\DFTUZ 98/05\\
    hep-th/9802076\\}}

\begin{center}
  {\Large {\bf Boundary dynamics and the statistical mechanics of the
      2+1 dimensional black hole}}
  
  \vspace{0.2in}
  
  M\'aximo Ba\~nados\footnote{On leave from Centro de Estudios
    Cient\'{\i}ficos de Santiago, Casilla 16443, Santiago, Chile and,
    Departamento de F\'{\i}sica, Universidad de Santiago, Casilla 307,
    Santiago, Chile.}  \vskip 0.2cm {{\it Departamento de F\'{\i}sica
      Te\'orica, Facultad de Ciencias,\\ Universidad de Zaragoza,
      Zaragoza 50009, Spain}
    \\
    \small E-mail: {\tt max@posta.unizar.es}} \vspace{0.2in}
  
  Thorsten Brotz \vskip 0.2cm {{\it Blackett Laboratory, Imperial
      College of Science, Technology
      and Medicine,\\
      Prince Consort Road, London SW7 2BZ, UK}
    \\
    \small E-mail: {\tt t.brotz@ic.ac.uk}} \vspace{0.2in}
  
  Miguel E. Ortiz \vskip 0.2cm {{\it Blackett Laboratory, Imperial
      College of Science, Technology
      and Medicine,\\
      Prince Consort Road, London SW7 2BZ, UK}
    \\
    \small E-mail: {\tt m.ortiz@ic.ac.uk}} \vskip 0.5truecm \today
\end{center}

\vskip 0.5truecm

\begin{abstract}
  We calculate the density of states of the 2+1 dimensional BTZ black
  hole in the micro- and grand-canonical ensembles. Our starting point
  is the relation between 2+1 dimensional quantum gravity and
  quantised Chern-Simons theory. In the micro-canonical ensemble, we
  find the Bekenstein--Hawking entropy by relating a Kac-Moody algebra
  of global gauge charges to a Virasoro algebra with a classical
  central charge via a twisted Sugawara construction.  This
  construction is valid at all values of the black hole radius.  At
  infinity it gives the asymptotic isometries of the black hole, and
  at the horizon it gives an explicit form for a set of deformations
  of the horizon whose algebra is the same Virasoro algebra.  In the
  grand-canonical ensemble we define the partition function by using a
  surface term at infinity that is compatible with fixing the
  temperature and angular velocity of the black hole.  We then compute
  the partition function directly in a boundary Wess-Zumino-Witten
  theory, and find that we obtain the correct result only after we
  include a source term at the horizon that induces a non-trivial
  spin-structure on the WZW partition function.
\end{abstract}

\medskip

\newpage

\section{Introduction}

The Chern-Simons formulation of 2+1 dimensional gravity \cite{AT-W}
has provided many interesting new insights into the problem of quantum
gravity.  (For a review see \cite{CarlipR}.)  Most notably, making use
of the relationship between 2+1 dimensional Chern-Simons theory and
the 1+1 dimensional WZW model \cite{Witten89,Moore-Seiberg}, Carlip
has argued for a statistical mechanical interpretation of the entropy
of a 2+1 dimensional black hole, backed up by a pair of tantalising
calculations which yielded the correct value for the black hole
entropy, both in Lorentzian \cite{Carlip1}, and Euclidean
\cite{Carlip2} formalisms.  Furthermore, the derivation given in
\cite{Carlip1} has been recently applied with success to de Sitter
space \cite{Maldacena-Strominger}, and may provide a tool for
understanding black hole entropy in string theory far from extremality
\cite{Sfetsos-Skenderis}.

The most important assumption in Carlip's analysis is that those
degrees of freedom responsible for the black hole entropy are located
at the horizon. This idea is certainly appealing and has been
advocated by many authors. However, at a technical and conceptual
level, it is difficult to see what states are counted in Carlip's
calculation, and what is being held fixed. In principle, it should be
possible to count states in a micro-canonical ensemble, holding the
mass and spin of the black hole fixed, or to infer the number of
states in a grand-canonical ensemble, holding fixed the black hole
temperature and angular velocity.

A different approach to understanding the statistical mechanical
origin of the 2+1 dimensional black hole entropy was recently proposed
by Strominger \cite{Strominger}.  In this approach, the basic
ingredient is the discovery by Brown and Henneaux \cite{bh} that the
asymptotic symmetry group of 2+1 dimensional gravity with a negative
cosmological constant \cite{dj} is the conformal group with a
(classical)
central charge
\begin{equation} 
c =\frac{3l}{2G},
\label{c} 
\end{equation} 
where $-1/l^2$ is the cosmological constant. Note that in the weak
coupling limit $G\rightarrow 0$, $c$ becomes very large.  Strominger
has pointed out that if one counts states by regarding the theory as
equivalent to $c$ free bosons, then at a fixed value of $L_0$ and
$\bar L_0$, the degeneracy of states gives rise to exactly the
Bekenstein--Hawking entropy.  Since in 2+1 dimensions $lM = L_0 + \bar
L_0$ and $J=L_0 - \bar L_0 $ where $M$ and $J$ are, respectively, the
black hole mass and angular momentum \cite{BHTZ}, Strominger's
computation is clearly a micro-canonical calculation. In this approach
one would like to find the underlying conformal field theory with a
central charge equal to (\ref{c}), and its connection is to a counting
of states at the black hole horizon. Since it is clear that one cannot
obtain the correct black hole entropy in general by looking only at
asymptotic isometries, it seems that Strominger's calculation succeeds
because the trivial nature of gravity in 2+1 dimensions results in an
isomorphism between boundary theories at infinity and at the horizon.

This paper has two goals. On the one hand, in Sec. 2 we present a
micro-canonical calculation of the black hole entropy. Starting with a
Chern-Simons theory, we find the algebra of global charges present at
any constant radius boundary surface in the black hole spacetime.  We
prove that a subset of this infinite set of conserved charges
satisfies the Virasoro algebra with a central charge equal to
(\ref{c}). This central charge, as in \cite{bh}, arises classically
\cite{max}. The advantage of considering the Chern-Simons formulation
is that the underlying conformal field theory is, at least at the
classical level, an $SL(2,R)\times SL(2,R)$ WZW model whose relation
to the Virasoro generators is via a twisted Sugawara construction
\cite{max}. Further, it is possible, to show that this Virasoro
algebra arises from a reduction of the WZW theory to a Liouville
theory \cite{henn}.  The counting of black hole microstates then
follows just as in \cite{Strominger}. As stressed before, since this
counting needs $M$ and $J$ fixed, the relevant states live in a
micro-canonical ensemble.

The subset of global charges satisfying the Virasoro algebra are shown
to be precisely those charges that, at infinity, leave the leading
order form of the metric invariant, agreeing with the asymptotic
isometries found in Ref. \cite{bh}.  However, the construction leading
to a Virasoro algebra of surface deformations is equally valid at any
radius, and in particular at the horizon. Thus, we are able to derive
the Bekenstein--Hawking entropy from the algebra of horizon
deformations.  We give the explicit form of the generators of the
Virasoro algebra and of the diffeomorphisms that they generate, valid
at a boundary surface located at any radius. 

Our second goal is to find the grand-canonical partition function for
the 2+1 dimensional black hole. In Sec. 3, we compute the partition
function $Z(\tau)$ for three dimensional Euclidean gravity on a solid
torus with a fixed value of the modular parameter $\tau$ of the torus.
We show that this parameter is related to the black hole inverse
temperature $\beta$ and angular velocity $\Omega$ by
\begin{equation} 
\tau = \hbar \beta
\left( \Omega + \frac{i}{l} \right),
\label{tau0} 
\end{equation} 
and therefore $Z(\tau)$ is clearly the grand-canonical partition
function. The computation of $Z(\tau)$ is not straightforward because
the relevant group is $SL(2,\C)$ which is not compact. We use here the
trick of replacing $SL(2,\C) \rightarrow SU(2) \times SU(2)$
\cite{Witten91,Carlip2} and show that the partition function correctly
accounts for the 2+1 dimensional black hole entropy, after continuing
the spin of the $SU(2)$ representations back to complex values. In
this case, we find the correct answer only if we include a source term
at the black hole horizon which has a particularly interesting
interpretation in terms of the WZW theory. It tracks the spin of each
representation and is the analogue of a $(-1)^F$ operator in fermionic
theories, twisting the WZW theory in the time direction.

\section{The micro-canonical ensemble}
\setcounter{equation}{0}

In this section we canonically quantise the degrees of freedom
associated with the gravitational field, by using the relation between
2+1 dimensional gravity and Chern-Simons theory. We then relate the
Kac-Moody algebra of the boundary WZW theory that emerges from the
Chern-Simons theory to a Virasoro algebra that describes asymptotic
isometries of the metric. This relation, achieved by a twisted
Sugawara construction, results in a theory whose degeneracy at fixed
mass and spin of the black hole leads to the Bekenstein--Hawking
entropy \cite{Strominger}.

\subsection{Global charges in Chern-Simons theory}

Let us begin with some general remarks on Chern-Simons theory, always
motivated by an application to the 2+1 dimensional black hole.
Consider the Chern-Simons action
\begin{equation}
  \label{a0}
  I_{CS}\left[A\right]={k\over 4\pi}\int
  \varepsilon^{\mu\nu\rho}\Tr \left(A_\mu\partial_\nu A_\rho + {2\over
      3} A_\mu A_\nu A_\rho\right)d^3x,
\end{equation}
where as we shall see below, we shall be interested in $k<0$. Up to a
boundary term, (\ref{a0}) can be put into the canonical form
\begin{equation}
  \label{a1}
  I[A_i,A_0]={k\over 4\pi} \int_{\Sigma\times R} 
\varepsilon^{ij} g_{ab} \left(
    A^a_i \dot A^b_j - A^a_t F^b_{ij}\right)dt d^2 x ,
\end{equation}
(here $\varepsilon_{t\rho\phi}=1$ and $\Tr(J_aJ_b)=g_{ab}$).
We take (\ref{a1}), without additional boundary terms, as our starting
point.  The variation of this action leads to
\begin{equation}
  \label{mm14a}
  \delta I\left[A\right]={k\over 4\pi}\int
  \varepsilon^{\mu\nu\rho}\Tr \left(\delta A_\mu
    F_{\nu\rho}\right)d^3x 
 -{k\over 2\pi}\int_{\pp\Sigma} \Tr A_t \, \delta A_\phi \, ,
\end{equation}
where we have assumed, in $t,\rho,\phi$ coordinates, that there is a
single outer boundary at fixed $\rho$. In order to make the variation
of this action well defined, we choose to fix one or other of the
conditions
\begin{equation}
A_t = \pm A_\phi
\label{chiral}
\end{equation}
at the boundary (in the $SL(2,\Re) \times SL(2,\Re)$ Chern-Simons
theory
we will choose $^\pm A_t = \pm ^\pm A_\pm$). However, this is not
enough to ensure the differentiability of (\ref{mm14a}) since under
(\ref{chiral}) the boundary term reduces to $\pm \delta
\int_{\pp\Sigma} \Tr A^2_\phi $. We thus need to imposse the equation
\begin{equation}
  \label{nn1}
  \pm \delta \int_{\pp\Sigma} \Tr A^2_\phi =0
\end{equation}
or, in other words, we need to fix the value of $\int \Tr A^2_\phi $
at the boundary. Under (\ref{chiral}) and (\ref{nn1}) the action has
well defined variations and the variational problem is then well
posed. We now analise the meaning of these boundary conditions.  

The chirality conditions (\ref{chiral}) should be regarded as
specifying the class of spacetimes or field configurations that will
be considered. In terms of lightlike coordinates $x^\pm = \phi \pm t$
these conditions read $A_\pm=0$ and it is well known that they are
satisfied by the BTZ black hole solutions (see, for example,
\cite{henn}).  These conditions leave a large residual symmetry group,
namely chiral (anti chiral) gauge transformations which are generated
by Kac-Moody fields, and the corresponding boundary degrees of freedom
are described by a chiral $WZW$ action. Since the black hole is
asymptotically anti-de Sitter, it is natural to further reduce the
boundary theory by imposing Polyakov's reduction conditions on the
$SL(2,\Re)$ currents leading to an effective Liouville theory, as done
in \cite{henn}. We shall mention this possibility in Sec. 2.8, but in
this paper we shall mainly consider the full Kac-Moody theory.  Let us
finally point out that the chirality boundary conditions
(\ref{chiral}) are not known to be in one to one correspondence with
the existence of a black hole, or an event horizon. The definition of
``black holes states" is a delicate problem which needs some
information about the topology as well as local fields.  Our strategy
here is to consider the chirality conditions, which include the black
hole solutions, as a starting point and quantise the theory on that
Hilbert space.  
 
Condition (\ref{nn1}) fixes the ensemble to be micro-canonical.
Indeed, in the $SL(2,\Re) \times SL(2,\Re)$ theory, the values of
$\int_{\pp\Sigma} \Tr\, ^\pm A^2_\phi$ are proportional to linear
combinations of the black hole mass and angular momenta. An
alternative procedure would be to add the term $\int_{\pp\Sigma} \Tr
A^2_\phi$ to the action (\ref{mm14a}) and leave its value
undetermined.  This corresponds to a canonical ensemble and will be
studied in detail in Sec. 3.

There will in general also be a boundary variation at any inner
boundary,
but in the following discussion we shall only consider charges on an
outer boundary.

In \cite{Balachandran,max} it was shown that the algebra of gauge
constraints leads to a set of global charges at the boundary whose
Poisson bracket algebra is a classical Kac-Moody algebra. These global
charges are equivalent to the global charges obtained by a reduction
of the Chern-Simons theory to a boundary WZW theory.

{From} the point of view of the three dimensional theory in
Hamiltonian
form, the global charges arise from considering the generators of
gauge transformations
\begin{equation}
  \label{a4}
  G(\eta^a)= -{k\over 4\pi}\int_\Sigma 
g_{ab}\eta^a \varepsilon^{ij}F^b_{ij} d^2x  +
Q(\eta^a) ,
\end{equation}
where $Q$ is a boundary term. Because of the presence of the boundary,
the functional derivative of $G(\eta)$ is well-defined only if the
boundary term $Q(\eta)$ has the variation
\begin{equation}
  \label{a5}
  \delta Q(\eta^a) = {k\over 2\pi}\int_{\pp \Sigma} g_{ab}\eta^a
\delta A^b_k dx^k.
\end{equation}
The Poisson bracket of two generators of the form (\ref{a4}) then
becomes,
\begin{equation}
  \label{a6}
  \left\{G(\eta),G(\lambda)\right\} = {k\over 4\pi}\int_\Sigma d^2 x
  f^a_{bc}\eta^b\lambda^c g_{ab} \varepsilon^{ij}F^b_{ij} + {k\over
    2\pi} \int_{\pp\Sigma} g_{ab} \eta^a D_k \lambda^b dx^k
\end{equation}
where $D_k \lambda^a = \partial_k \lambda^a + f^a_{bc} A^b_k
\lambda^c$.
One expects the boundary term in the right hand side of (\ref{a6}) to
be equal to the charge $Q(f^a_{bc}\eta^b\lambda^c)$, plus a possible
central term \cite{bh2}. But to check this we first need to give
boundary conditions in order to integrate (\ref{a5}) and extract the
value of $Q$. We shall consider two different classes of boundary
conditions.

\subsubsection{Gauge charges}

Assuming that $\eta^a$ is fixed at the boundary the charge is
\begin{equation}
  \label{a7}
  Q(\eta) = {k\over 2\pi}\int g_{ab} \eta^a A^b_k dx^k.
\end{equation}
One can then check that, indeed, the boundary term in (\ref{a6})
contains the charge associated to the commutator $[\eta,\lambda]$ plus
a central term.  Imposing the constraints, the algebra of global
charges becomes
\begin{equation}
  \label{a8}
  \left\{Q(\eta),Q(\lambda)\right\}_{DB} = -Q([\lambda,\eta]) +
  {k\over
    2\pi}\int g_{ab} \eta^a \partial_k \lambda^b dx^k.
\end{equation}
Since we are interested in the Dirac bracket algebra, we should not
only solve the constraint on the bulk but also fix a gauge. It is
convenient to use the gauge chosen in \cite{max}, and in \cite{henn}
in a parallel treatment using the WZW formulation,
\begin{equation}
  \label{a8a}
  A_\rho = b(\rho)^{-1}   \partial_\rho b(\rho),
\end{equation}
where the boundary is taken to be at fixed $\rho$. This
gauge choice together with the constraint $F^a_{r\varphi}=0$ imply
\begin{equation}
 A_\phi = b(\rho)^{-1} A(\phi,t) b(\rho).
\label{}
\end{equation}
The gauge choice (\ref{a8a}) is preserved only by gauge
transformations whose parameters are of the form
\begin{equation}
  \label{a8aa}
  \eta(\rho,\phi,t) = b^{-1}(\rho)\lambda(\phi,t)b(\rho).
\end{equation}
Since $\eta$ still contains an arbitrary function of time
$\lambda(\phi,t)$, it seems that the gauge freedom has not been fixed
completely yet.  The extra requirement that fixes the time dependence
of the gauge parameters comes from the boundary condition on $A_\mu$.
We have chosen our action in order to impose one or other of the
conditions
\begin{equation}
  \label{aa8}
  A_t = \pm A_\phi.
\end{equation}
These conditions remove the gauge invariance since the group of
transformations leaving (\ref{aa8}) invariant does not contain any
arbitrary function of time.  $\lambda$ is constrained to depend only
on $t+\phi$ or $t-\phi$. Setting,
\begin{equation}
  \label{a10}
  A^a(\phi,t) = -{1\over k}\sum_n g^{ab} T_{b\ n}(t) e^{- in\phi},
\end{equation}
equation (\ref{a8}) leads to the classical Kac-Moody algebra
\begin{equation}
  \label{a11}
  \left\{T_{a\ m},T_{b\ n}\right\} = f^c_{ab}T_{c\ m+n} - ik m g_{ab}.
  \delta_{m+n}
\end{equation}
Note that the central term has the usual sign for $k<0$.

We could have obtained the same algebra by inserting (\ref{a8a}) into
the action (\ref{a1}) and computing the resulting Poisson brackets
(see the next section for a detailed discussion of this reduction).

\subsubsection{Diffeomorphisms}

We have seen above that global charges associated to gauge
transformations that do not vanish at the boundary give rise to an
infinite number of conserved charges satisfying the Kac-Moody algebra.
We shall now investigate those charges associated to the group of
diffeomorphisms at the boundary. Since the boundary is a circle, one
expects to find the Virasoro algebra. Furthermore, since in
Chern-Simons theory diffeomorphisms and gauge transformations are
related, one expects the Virasoro and Kac-Moody generators to be
related by the Sugawara construction. Actually, for those
diffeomorphisms with a non-zero component normal to the boundary, one
finds a twisted Sugawara construction which induces a classical
central charge in the Virasoro algebra \cite{max}.  This central
charge was first found in \cite{bh} in the ADM formulation of 2+1
dimensional gravity and has recently been shown to play an important
role in understanding the statistical mechanical origin of the 2+1
dimensional black hole entropy \cite{Strominger}.

Recall that in Chern-Simons theories, diffeomorphisms with parameter
$\xi^\mu$ are related to gauge transformations with parameter
$\eta^a=A^a_\mu \xi^\mu$ by the equations of motion.  In a canonical
realisation of gauge symmetries, $A^a_t$ is a Lagrange multiplier, and
so we must instead consider gauge transformations
\begin{equation}
  \label{mm1b}
  \eta^a =  \xi^i A^a_i.
\end{equation}
As we did in the last section, we fix the gauge by fixing $A_\rho =
b^{-1} \pp_\rho b$. We also choose coordinates for the on-shell
solution for which $b=e^{\rho\alpha}$ which implies
\begin{equation}
  \label{a133}
  A^a_\rho = \alpha^a.
\end{equation}
This will be a good choice of the radial coordinate at infinity for
the black hole. As a consequence of this choice, diffeomorphisms
$\xi^i$ that preserve the gauge choice (\ref{a8a}) and (\ref{a8aa})
must be independent of $\rho$. Since the gauge choice only fixes the
gauge freedom in the interior of the manifold, we are able to derive
the algebra of global boundary diffeomorphisms in complete generality
by looking only at $\xi^i(\phi,t)$, subject to the constraints imposed
by the boundary condition on the gauge field.

Since from (\ref{mm1b}) the gauge parameter is field-dependent, we
replace (\ref{a7}) by
\begin{equation}
  \label{a14}
  Q(\xi) = -{k\over 4\pi}\int g_{ab} \left( 2\xi^r \alpha^a A^b + 
    \xi^\phi A^a A^b \right)d\phi. 
\end{equation}
This is a good choice of $Q(\xi)$ since $A_\rho$ is left unchanged by
the action of the global charges. In other words $A_\rho=\alpha$ is
fixed at the boundary.  The algebra of charges then becomes \cite{max}
\begin{equation}
  \label{a15}
\left\{Q(\xi),Q(\zeta)\right\}_{DB} = - Q([\xi,\zeta]) + {k\over
2\pi}g_{ab}\alpha^a\alpha^b\int \xi^r\partial_\phi\zeta^r \:  d\phi .
\end{equation}
If we restrict the diffeomorphisms to be of the specific form
\cite{max,carlipwhat} (see below for a geometrical justification
for this restriction),
\begin{equation}
  \label{a16}
  \xi^i=(-\beta \partial_\phi \xi,\xi),
\end{equation}
then the algebra of these restricted diffeomorphisms is the continuous
form of the Virasoro algebra with central charge
\begin{equation}
  \label{a18}
  \left\{Q(\xi),Q(\zeta)\right\}_{DB} = -Q([\xi,\zeta]) - {k\over
  2\pi}
  \alpha^2 \beta^2 \int \xi\partial^3_\phi \zeta \: d\phi
\end{equation}
where $\alpha^2$ denotes $\alpha^a \alpha^b g_{ab}$.

Defining
\begin{equation}
  \label{a19}
  \sum_n L_n e^{-in\phi} = -{k\over 2}g_{ab}\left(\alpha^a\alpha^b
  \beta^2 +
    2\alpha^a\partial_\phi A^b + A^a A^b\right),
\end{equation}
or equivalently
\begin{equation}
  \label{a19b}
 L_n = -{1\over 2k}\sum_m g^{ab} T_{a\ m} T_{b\ n-m} - i n
  \alpha^a
  T_{a\ n} - {k\over 2}\alpha^2 \beta^2 \delta_n,
\end{equation}
gives the usual Poisson bracket version of the Virasoro 
algebra
\begin{equation}
  \label{a19c}
  \left\{L_m,L_n\right\} = i(m-n)L_{m+n} - ik\alpha^2\beta^2
m(m^2-1)\delta_{m+n} ,
\end{equation}
so that the central charge is
\begin{equation}
  \label{a19ca}
  c={-12k\alpha^2 \beta^2},
\end{equation}
which is positive for $k<0$\footnote{ Note that the appearance of 
  the term proportional to $m$ in the Virasoro algebra is due to the 
  shift of the $L_0$ operator by $ - {k\over 2} g_{ab} \alpha^a
  \alpha^b \beta^2$ in (\ref{a19b}).}. 
Hence, as expected, those
diffeomorphisms that lead to global charges, after the restriction
(\ref{a16}), induce an infinite number of conserved charges satisfying
the Virasoro algebra with a classical central charge.  Eq.
(\ref{a19b}) is an example of a twisted Sugawara construction
\cite{go}. Note that from (\ref{a19}) we see that the boundary
condition (\ref{nn1}) fixes the value of $L_0$ at the boundary.

\subsection{2+1 dimensional Chern-Simons gravity}

In 2+1 dimensional gravity with a negative cosmological constant, 
the Einstein--Hilbert action is represented by
the difference of two Chern-Simons actions
\begin{equation}
I_{\rm CS} = I\left[\A\right]-I\left[\AA\right],
\label{mm13}
\end{equation}
for a pair of $SL(2,R)$ gauge fields $\A$ and $\AA$, where\footnote{We
  take $J_0 = {1\over 2}\left(\matrix{0 & -1 \cr 1 & 0}\right)$, $J_1
  = {1\over 2}\left(\matrix{1 & 0 \cr 0 & -1}\right)$, $J_2 = {1\over
    2}\left(\matrix{0 & 1 \cr 1 & 0}\right)$ so that $[J_a,J_b] =
  {\varepsilon^c}_{ab}J_c$, and $\Tr \left(J_aJ_b\right) = {1\over
    2}\eta_{ab}$ where $\varepsilon_{012}=1$ and $\eta_{ab}={\rm
    diag}(-1,1,1)$.}
\begin{equation}
I\left[^{(\pm)}A\right]={k\over 4\pi}\int \varepsilon^{ij}\Tr 
\left(^{(\pm)}A_i
  {^{(\pm)}\dot{A}_j}-^{(\pm)}A_t \,
^{(\pm)}F_{ij}\right)d^3x \, .
\label{mm14}
\end{equation}
The Einstein--Hilbert action is recovered by defining
\begin{equation}
{^{(\pm)} A}^a_\mu = \omega^a_\mu \pm {e^a_\mu\over l},
\label{mm15}
\end{equation}
from which it follows that
\begin{equation}
I_{\rm CS}={k\over 4\pi l}\int
\sqrt{-g}\left(R + {2\over l^2}\right) \, d^3x\,, 
\label{mm17}
\end{equation}
ignoring boundary terms. This relates the level $k$ of the
Chern-Simons theories to Newton's constant,
\begin{equation}
k=-{l\over 4G}.
\label{mm18}
\end{equation}
We see that for the black hole, $k<0$, explaining why we have
developed our arguments for negative $k$.

\subsection{Diffeomorphisms and gauge transformations in 2+1
  dimensional Chern-Simons gravity}

In the gauge theory representation of 2+1 dimensional gravity with a
negative cosmological constant, it is possible to reproduce the full
diffeomorphism transformation properties of $e^a_\mu$ and
$\omega^a_\mu$ by a gauge transformation in both the covariant and
canonical formalisms.  In the covariant formalism, this gauge
transformation must be chosen so that the gauge parameters for the
connections ${^{(\pm)} A}^a_\mu$ are equal to
\begin{equation}
  \label{mm1a}
  ^{(\pm)} \lambda^a =  \xi^\mu \, {^{(\pm)} A}^a_\mu,
\end{equation}
where $\xi^\mu$ is the same in both cases.

In the canonical formalism, the situation is a little more
complicated, and perhaps not well-known, so we devote some space to
it.  If in this case we set ${^{(+)}\xi}^i={^{(-)}\xi}^i$, then
there are only two arbitrary functions that parametrise
diffeomorphisms, and it is easy to see that these two parameters only
generate spatial diffeomorphisms. We are thus led to consider the case
${^{(+)}\xi}^i{\ne}{^{(-)}\xi}^i$.

In a canonical theory, diffeomorphisms and Lorentz transformations are
realised by the action of the constraints. In this case, the
Hamiltonian is equal to
\begin{equation}
  \label{mm18c}
  H = {1\over 16\pi G}
  \int d^2x \left[e^a_t\varepsilon^{ij}\left(R_{aij} + {1\over
        l^2}\varepsilon_{abc} e^b_i e^c_j\right) +
    \omega^a_t\left(D_i e_{aj}
      -D_je_{ai}\right) \right]\, ,
\end{equation}
Via the two Lagrange multipliers $e^a_t$ and $\omega^a_t$, the
Hamiltonian induces a diffeomorphism defined by $e^a_t= \chi^\perp n^a
+ \chi^k e^a_k$ and a Lorentz transformation defined by $j^a =
\omega^a_t$. Explicitly,
\begin{eqnarray}
\delta_{(\chi^\mu,j^a)} e^a_i &=& \partial_i \left(\chi^\perp n^a + 
\chi^k e^a_k\right)
+ {\varepsilon^a}_{bc}\omega^b_i \left(\chi^\perp n^c + \chi^k
  e^c_k\right) - {\varepsilon^a}_{bc}j^b e^c_i , \label{mm5} \\
\delta_{(\chi^\mu,j^a)} \omega^a_i &=& 
{1\over l^2}{\varepsilon^a}_{bc}e^b_i \left(\chi^\perp n^c + \chi^k
  e^c_k\right) + \partial_i j^a + {\varepsilon^a}_{bc} \omega^b_i j^c.
\label{mm5a}
\end{eqnarray}

Let us now compare these transformations laws with those obtained by a
gauge transformations parametrised by $^{(\pm)}\eta^a =
{^{(\pm)}\xi}^i \,{^{(\pm)}A}^a_i$. It is convenient to define
\begin{equation}
V^i={1\over 2}\left(^{(+)}\xi^i + ^{(-)}\xi^i\right),\qquad
W^i={l\over 2}\left(^{(+)}\xi^i - ^{(-)}\xi^i\right).
\label{mm8}
\end{equation}
The transformation equations for $e^a_i$ and $\omega^a_i$ read
\begin{eqnarray}
\delta_{(V^i,W^i)} e^a_i &=& \partial_i 
\left(e^a_j V^j + \omega^a_j W^j\right)
+ {\varepsilon^a}_{bc}\omega^b_i\left(e^c_j V^j + \omega^c_j
W^j\right) -
{1\over l^2}{\varepsilon^a}_{bc}
\left(e^b_j  W^j\right)e^c_i ,
\\
  \delta_{(V^i,W^i)} \omega^a_i &=&  
{1\over l^2}\left[\partial_i\left(e^a_j W^j\right) +
{\varepsilon^a}_{bc} \omega^b_i \left(e^c_j W^j\right)
+{\varepsilon^a}_{bc} e^b_i\left(e^c_j V^j +
\omega^c_jW^j\right)\right].
\label{40}
\end{eqnarray}
Comparing this with (\ref{mm5}) and (\ref{mm5a}) we recognise these
expressions as the canonical formulae for diffeomorphisms parametrised
by
\begin{equation}
  \label{mm10b}
  \chi^\perp n^a + \chi^i e^a_i = e^a_i V^i + \omega^a_i W^i,
\end{equation}
along with a Lorentz transformation with parameter
$j^a=e^a_j{W}^j/l^2$.  Using
\begin{equation}
  \label{mm10d}
  n^a = -{1\over
  2\sqrt{h}}{\varepsilon^a}_{bc}e^b_ie^c_j\varepsilon^{ij},
\end{equation}
($h_{ij}=e^a_i e_{aj}$ and is used to raise and lower spatial
indices), we can pick out
\begin{equation}
  \label{mm10e}
  \chi^\perp = {1\over 2\sqrt{h}}
  {\varepsilon}_{abc}\omega^a_ie^b_je^c_kW^i\varepsilon^{jk},
\end{equation}
and
\begin{equation}
  \label{mm10f}
  \chi^i = V^i + e^i_a \omega^a_j W^j.
\end{equation}
We can now see
explicitly that if we had set $W^i=0$, then the gauge transformations
(\ref{40}) would not generate timelike diffeomorphisms.

\subsection{The 2+1 dimensional black hole}

The classical black hole solution \cite{BTZ} in Lorentzian signature
can be conveniently written in proper radial coordinates as
\begin{equation}
ds^2 = -\sinh^2\rho\left({r_+ dt\over l}+r_- d\phi\right)^2 + 
l^2d\rho^2 +\cosh^2\rho\left({r_-
dt\over l} + r_+ d\phi\right)^2.
\label{mm19}
\end{equation}
In these coordinates, the horizon is at $\rho=0$. $\phi$
is an angular coordinate with period $2\pi$. Note that the above
metric represents only the exterior of the black hole.  The inner
regions can be obtained by replacing some hyperbolic functions by
their trigonometric partners. The mass $M$ and angular momentum $J$ of
the black hole are given in terms of $r_\pm$ as
\begin{equation}
  \label{mm19a}
  M={r_+^2 + r_-^2\over 8Gl^2},\qquad J={2r_+r_-\over 8Gl},
\end{equation}
and the relation between the Schwarzschild radial coordinate $r$ and
the proper radial coordinate $\rho$ is
\begin{equation}
r^2 = r_+^2 \cosh^2\rho -r_-^2 \sinh^2\rho.
\label{rho}
\end{equation}

By going to its Euclidean section (see Eq. (\ref{mm19eu}) below), the
black hole (\ref{mm19}) can be seen to have a temperature
\begin{equation}
  \label{Temp}
  T={\hbar\left(r_+^2-r_-^2\right)\over 2\pi l^2 r_+},
\end{equation}
and, using the first law of black hole mechanics, $\delta M = T \delta
S + \Omega \delta J$, we find an entropy
\begin{equation}
  \label{Ent}
  S={2\pi r_+\over 4\hbar G}.
\end{equation}

The metric can be written in first order form as
\begin{eqnarray}
e^0 &=& \left({r_+ dt\over l} + r_- d\phi\right) \sinh\rho ,
\cr
e^1 &=& l d\rho,
\cr
e^2 &=& \left( {r_- dt\over l} + r_+ d\phi\right)\cosh\rho,
\label{mm20}
\end{eqnarray}
so that after computing the spin connection, the gauge connection 
representing the black hole is given by
\begin{eqnarray}
^{(\pm)} A^0 &=& \pm {r_+\pm r_-\over l}\left({dt\over l}\pm
d\phi\right)
\sinh\rho ,
\cr
^{(\pm)} A^1 &=& \pm d\rho,
\cr
^{(\pm)} A^2 &=& {r_+\pm r_-\over l}\left({dt\over l}\pm d\phi\right)
\cosh\rho,
\label{mm22}
\end{eqnarray}
or in matrix form,
\begin{equation} 
^{(\pm)} A = {1\over 2}\left(\matrix{
\pm d\rho & z_\pm e^{\mp
  \rho} 
dx^\pm 
\cr
z_\pm e^{\pm \rho} 
dx^\pm & \mp d\rho}\right),
\label{mm23}
\end{equation}
where $x^\pm = t/l\pm \phi$ and $z_\pm = (r_+\pm r_-)/l$.

We can put this solution into the gauge (\ref{a8a}),
\begin{equation}
^{(\pm)} \alpha = \pm J_1,\qquad {^{(\pm)}b} =
\exp\left({^{(\pm)}\alpha}\rho\right)  = 
\left(\matrix{e^{\pm\rho/2} & 0\cr
0& e^{\mp \rho/2}}\right),
\label{mm24}
\end{equation}
so that
\begin{equation}
^{(\pm)} A = \pm z_\pm J_2.
\label{mm25}
\end{equation}
We then see that for the black hole
\begin{equation}
\alpha^2=1/2.
\label{alpha2}
\end{equation}
(Note that this value of $\alpha^2$ can be changed by a rescaling
  of $\rho$.)

We can see from (\ref{mm25}) that the gauge connection $A_\phi$ leads
to a non-trivial holonomy around the closed loops of constant $\rho$
and $t$,
\begin{equation}
  \label{hol}
  \Tr{\cal P}\exp\oint {^{(\pm)}A}_\phi d\phi = 
  2\cosh \left(\pi z_\pm\right) = 2\cosh\left(\pi\sqrt{8G(M\pm
      J/l)}\right).
\end{equation}

\subsection{Global charges and the 2+1 dimensional black hole}

The first step in discussing the algebra of global charges for the 2+1
dimensional black hole is to choose appropriate boundary conditions
for the gauge fields ${^{(\pm)}A}^a_\mu$. From the on-shell gauge
fields (\ref{mm22}), it is natural to impose the conditions
\begin{equation}
  \label{mm222}
  \A^a_t = \A^a_\phi,\qquad \AA^a_t = -\AA^a_\phi,
\end{equation}
which lead to the conditions 
\begin{equation}
  \label{m223}
  \partial_\mp{^{(\pm)}\xi}^i = 0
\end{equation}
on the diffeomorphism parameters.  This, along with the constraint
\begin{equation}
  \label{27ab}
  \partial_\rho{^{(\pm)}\xi}^\phi=0,
\end{equation}
then defines the complete set of diffeomorphisms that leave the
boundary conditions invariant and preserve the gauge (\ref{a8a}) and
(\ref{a8aa}).

We have seen above that if we impose on these diffeomorphisms the
supplementary condition
\begin{equation}
  \label{mm25cc}
  {^{(\pm)}\xi}^\rho = -\beta \partial_\phi{^{(\pm)}\xi}^\phi ,
\end{equation}
then the algebra of global charges associated with the remaining
diffeomorphisms,
\begin{equation}
  \label{dif}
  {^{(+)}\xi}^i =
  \left(-\beta
  \partial_\phi{^{(+)}\xi}^\phi(x^+),{^{(+)}\xi}^\phi(x^+) 
  \right),\qquad 
  {^{(-)}\xi}^i = 
  \left(-\beta
  \partial_\phi{^{(-)}\xi}^\phi(x^-),{^{(-)}\xi}^\phi(x^-)
  \right)
\end{equation}
leads to a pair of Virasoro algebras, one sector coming from each of
the gauge
fields, with central charge $c=-12k\alpha^2\beta^2$.

In terms of the gauge field (\ref{mm22}) representing the black hole,
the global charges (without condition (\ref{mm25cc})) generate the
transformations
\begin{eqnarray} 
\delta\, ^{(\pm)} A_\phi
&=& {1\over 2}\left(\matrix{
\mp \partial_\phi{^{(\pm)}\xi}^\rho & 
z_\pm\displaystyle e^{\mp\rho} 
\left({^{(\pm)}\xi}^\rho \mp \partial_\phi{^{(\pm)}\xi}^\phi\right)
\cr
z_\pm\displaystyle e^{\pm \rho} 
\left(-{^{(\pm)}\xi}^\rho \mp\partial_\phi{^{(\pm)}\xi}^\phi\right)& 
\pm \partial_\phi{^{(\pm)}\xi}^\rho }\right),
\cr
\delta\,^{(\pm)} A_\rho &=& 0.
\label{mm27aa}
\end{eqnarray}
Let us now look at the form of these transformations at infinity
(corresponding to placing the boundary at infinity). Then, focusing on
the leading order terms of order $e^\rho$, we find that conditions
(\ref{mm25cc})  with $\beta =1$ are precisely what is required for
them to vanish. It is easy to check that rescaling the coordinate
$\varrho$ (this means changing the value $\alpha^2$) introduces the
(more general) condition 
\begin{equation} \label{condition}
 \alpha^2 \beta^2 =\frac{1}{2},
\end{equation}
which of course includes the above discussion of the black hole 
solution in the gauge (\ref{mm22}).
Thus the sub-algebra of global charges defined by (\ref{dif}) 
and the condition (\ref{condition}) generates
asymptotic isometries of the black hole metric. Note that these
automatically include the anti-de Sitter group $SO(2,2)$.

Let us make a direct comparison with the asymptotic isometries found
in \cite{bh}. Using the results of the last subsection, we can
translate the action of the transformations (\ref{dif}) into the
action of a temporal and spatial diffeomorphism. Using the on-shell
values of $\omega^a_i$ and $N$, we see from (\ref{mm10e}) and
(\ref{mm10f}), and from the appropriate coordinate relations that
\begin{eqnarray}
  \label{mm25ad}
  \chi^\perp &=& N\, {^{(3)}\chi}^t = N W^\phi ,
\cr
\chi^i &=& {^{(3)}\chi}^i + N^i\, {^{(3)}\chi}^t = V^i + N^i W^\phi,
\end{eqnarray}
so that we get the diffeomorphism
\begin{eqnarray}
  \label{mm25ae}
  {^{(3)}\chi}^t &=& {l\over 2}\left({^{(+)}\xi}^\phi(x^+) - 
    {^{(-)}\xi}^\phi(x^-)
  \right)
  \cr
  {^{(3)}\chi}^\rho &=& -{1\over 2}\left(
    \partial_\phi{^{(+)}\xi}^\phi(x^+) +
    \partial_\phi{^{(-)}\xi}^\phi(x^-)\right) 
  \cr
  {^{(3)}\chi}^\phi &=& {1\over 2}\left({^{(+)}\xi}^\phi(x^+) +
    {^{(-)}\xi}^\phi(x^-)\right)
\end{eqnarray}
accompanied by a Lorentz transformation with parameter ${e^a_i
  {W}^i/l}$. Comparing (\ref{mm25ae}) with the asymptotic isometries
found in Ref. \cite{bh}, 
\begin{eqnarray}
  \label{mm25a}
  {^{(3)}\chi}^t &=& l\left(T^+(x^+) +
  T^-(x^-)\right)+{l^3e^{-2\rho}\over 2}
  \left(\partial^2_+ T^+ + \partial^2_-T^-\right) + {\cal
    O}(e^{-4\rho}),
  \cr
  {^{(3)}\chi}^\rho &=& -\left(\partial_+ T^+ + \partial_- T^-\right)
  + {\cal
    O}(e^{-\rho}),
  \cr
  {^{(3)}\chi}^\phi &=& T^+ - T^- -{l^2e^{-2\rho}\over 2}
  \left(\partial^2_+ T^+ - \partial^2_-T^-\right)+ {\cal
    O}(e^{-4\rho}),
\end{eqnarray}
we see that there is exact agreement to leading order. (Here
$\partial_\pm = (l\partial/\partial_t\pm\partial/\partial\phi)/2$, and
note that $\partial_\pm T^\pm = \pm\partial_\phi T^\pm$.)  The
disagreement to sub-leading order is because the diffeomorphisms
(\ref{mm25a}) do not preserve our gauge choice in the interior. It
seems that up to a choice of gauge in the interior  (which is
irrelevant since we are trying to isolate the boundary dynamics),
(\ref{mm25a}) and (\ref{dif}) are equivalent.

We know from the analysis of global charges that they lead to a
Virasoro algebra with central charge $c=-12k\alpha^2 \beta^2$.
Inserting (\ref{condition})  and the value of $k$
given by (\ref{mm18}), we find that
\begin{equation}
  \label{mm25cd}
  c={3l\over 2G},
\end{equation}
which agrees with the result obtained in \cite{bh} for the algebra of
asymptotic isometries. As a result, we see that the algebra of
diffeomorphisms obtained by Brown and Henneaux is related to the
Kac-Moody algebra of the boundary WZW theory at infinity by the
twisted Sugawara construction (\ref{a19b}).

\subsection{A Virasoro algebra at all $\rho$}

Perhaps the most important point about the analysis of global charges
is that it goes through for a boundary located on any surface of
constant $\rho$. Thus the set of diffeomorphisms defined by 
  (\ref{dif}) and (\ref{condition}) leads to a Virasoro algebra of
  global charges with $c=-6k$ on any such boundary. However, the
  connection between the Virasoro algebra and isometries of the three
  metric appears to be valid only at infinity. At finite $\rho$, the
  Virasoro algebra is a subset of all deformations of that surface,
  but without any obvious property to distinguish it.  In particular,
  if we take the boundary to be at the horizon, we find that the
  global charges that generate the Virasoro algebra generate
  a particular subset of deformations of the horizon with components
  both tangential and normal to the horizon, described by
  (\ref{mm25ad}) and (\ref{mm25ae}). We cannot, of course, rule out
  the possibility that these deformations may have some distinguishing
  properties that remain undiscovered.

If at finite $\rho$, the subset of global charges which give rise
to the Virasoro algebra are not special in any way, perhaps one should
consider all generators
\begin{equation}
  \label{gen}
  {^{(\pm)}\xi}^i(x^\pm)
\end{equation}
on an equal footing, and regard the condition (\ref{mm25cc}) as a
technical step that leads to the Virasoro algebra. The fact that the
Virasoro algebra is a subalgebra of the algebra of deformations then
suggests that the number of states generated by the larger algebra
should be greater than or equal to the Bekenstein value. We shall
discuss this issue briefly in the conclusions.

\subsection{Density of states}

We have so far in this section derived the algebra of global charges
on any boundary of fixed $\rho$, and shown that a subset of them leads
to the Virasoro algebra with a classical central term. We have also
made explicit the relation between the asymptotic isometries of Ref.
\cite{bh} and this subset of global charges when defined at infinity.
We may now count states in the conformal field theory, by looking at
representations of the Virasoro algebra, as done in \cite{Strominger}.
We must look for representations with a specific value of $L_0$ and
$\bar{L}_0$, since according to (\ref{a19b}), these two quantities are
related to the mass and spin of the black hole as
\begin{eqnarray}
  \label{m224}
  L_0 &=& -{k\over 4\pi}\int\left(g_{ab}\, \A^a 
    \A^b + \frac{1}{2} \right)d\phi = {1\over 2}\left(lM+J\right) +
  {l\over 16G},
\cr
  \bar{L}_0 &=& -{k\over 4\pi}\int\left(g_{ab}\, \AA^a 
    \AA^b + \frac{1}{2} \right)d\phi ={1\over 2}\left(lM-J\right) +
  {l\over 16G},
\end{eqnarray}
or, neglecting the $l/16G$ terms,
\begin{equation}
  \label{m225}
  M={L_0+\bar{L}_0\over l},\qquad J=L_0-\bar{L}_0.
\end{equation}
Note that fixing (\ref{nn1}) is equivalent to fixing $M$ and
$J$, justifying our choice of action (\ref{a1}) for the
micro-canonical ensemble.

As pointed out in \cite{Strominger}, since $c$ is large, one can use
the degeneracy formula for $c$ free bosons to deduce a density of
states equal to
\begin{equation}
  \label{mm1111}
  \rho(M,J) = \exp\left(2\pi r_+\over 4\hbar G\right),
\end{equation}
which agrees with the Bekenstein--Hawking entropy of the 2+1
dimensional black hole.

It is interesting that now this analysis is not necessarily related to
diffeomorphisms at infinity. We can think of these states as living on
any surface of constant $\rho$, and in particular they could be
defined at the horizon. As far as we are aware, this system then
provides the first explicit realisation of a set of deformations of a
black hole horizon that can be quantised to yield the correct
Bekenstein--Hawking entropy.  

\subsection{A Liouville action for the Virasoro algebra}

We end this section with a brief remark about the relation between the
Virasoro algebra (\ref{a19c}) and the reduction of WZW theory to
Liouville theory as discussed in \cite{henn} in the context of 2+1
dimensional gravity, and by a number of other authors in a more
general context (see \cite{raf} for an extensive list of references).
As explained in \cite{henn}, the first step is to join the two chiral
WZW theories into a single non-chiral WZW theory. Then the reduction
takes place by imposing certain constraints on the currents of the WZW
theory and interpreting a second set of constraints as gauge fixing
conditions.

Referring back to the conditions (\ref{mm25cc}) and (\ref{m223}), we
see that (\ref{m223}) are just the chirality conditions for each
$SL(2,R)$ sector. As we shall see below, in the WZW theories, these
conditions arise from the dynamics of the WZW currents.  Eqs.
(\ref{mm25cc}) are equivalent to holding fixed
\begin{equation}
  \label{fix}
  {1\over 2}\left(
    {^{(\pm)}A}^0_\phi \pm {^{(\pm)}A}^2_\phi\right) = z_\pm,
\end{equation}
and are equivalent to the constraints usually imposed in the reduction
from WZW to Liouville theory \cite{henn,raf}. The reduction is
completed by a set of gauge fixing conditions on the currents. A
direct application of the results of \cite{henn} uses the simplest
gauge fixing condition, ${^{(\pm)}A}^3_\phi=0$. Looking at the set of
diffeomorphisms (\ref{a16}) that lead to (\ref{a19c}), one can see
from (\ref{mm27aa}) that although ${^{(\pm)}A}^3_\phi=0$ on-shell,
\begin{equation}
  \label{lio2}
  \delta\,{^{(\pm)}A}^3_\phi=\pm \partial_\phi^2{^{(\pm)}\xi}^\phi.
\end{equation}
Thus, Dirac brackets will be required to compute the operator algebra
for the Liouville theory.

We can invoke Ref. \cite{raf} and see that for any
gauge fixing condition, the constraints (\ref{mm25cc}) lead to a
Liouville theory with a central charge that for large $k$ is equal to
\cite{ack}
\begin{equation}
  \label{cc}
  c=-6k,
\end{equation}
(since we must use the non-standard convention that has $k<0$).  This
agrees with the result we have obtained for the central charge
from
(\ref{a19ca}).

We conclude that the Virasoro algebra (\ref{a19c}) can be interpreted
as coming from an underlying Liouville theory, as predicted in Refs.
\cite{henn} and \cite{Strominger}.

\section{The grand-canonical partition function}
\setcounter{equation}{0}

The goal of this section is to compute the grand-canonical partition
function for three dimensional gravity
\begin{equation}
Z(\beta,\Omega) = \int D[e]D[\omega] 
\exp\left( -{1\over \hbar}I_{EH}[e,\omega;\beta,\Omega]\right)
\end{equation}
where $\beta$ and $\Omega$ are, respectively, the inverse temperature
and angular velocity of the black hole. These two (intensive)
parameters define the grand-canonical ensemble. The thermodynamic
quantities such as average energy and entropy can then be obtained
from $Z$ through the thermodynamic formulae
\begin{eqnarray}
\left\langle E\right\rangle 
&=& \frac{\Omega}{\beta} \frac{\pp \log Z}{\pp \Omega} - 
\frac{\pp \log Z}{\pp \beta}, \\ 
\left\langle J\right\rangle 
&=& \frac{1}{\beta} \frac{\pp \log Z}{\pp \Omega}, \\ 
S &=& \log Z - \beta \frac{\pp \log Z}{\pp \beta},
\label{S}
\end{eqnarray}
since
\begin{equation}
  \label{max2}
  Z(\beta,\Omega) = \int dE dJ \rho(E,J) e^{-\beta E - \beta\Omega J}
  = 
  e^{-\beta\langle E\rangle -\beta\Omega\langle J\rangle + S}.
\end{equation}

The problem now requires two steps: First, we need to impose boundary
conditions in the action principle such that the action has well
defined variations for $\beta$ and $\Omega$ fixed, and second, we need
to actually compute $Z(\beta,\Omega)$.

\subsection{Euclidean three dimensional gravity}

The grand-canonical partition function will be defined as a sum over
Euclidean metrics. The Einstein--Hilbert action for Euclidean gravity
with a negative cosmological constant may again be represented by the
difference of two Chern--Simons actions, but now for the group
$SL(2,\C)$ \cite{AT-W}. We shall use this property to compute the
partition function.

Defining,
\begin{equation}
A^a=w^a+\frac{i}{l}e^a, \spc \bar A^a = w^a - \frac{i}{l} e^a,
\label{A}
\end{equation}
and\footnote{Our conventions are $[J_a,J_b] = \epsilon_{abc} J^c$ and
  $\Tr(J_aJ_b) = -(1/2) \delta_{ab}$.} $A = A^a J_a$, $\bar A = \bar
A^a J_a$, then up to boundary terms,
\begin{equation}
I_{EH} = \frac{1}{16\pi G} \int_M \sqrt{g} \left(R +
\frac{2}{l^2}\right)
=i\left(I\left[A\right]- I\left[\bar A\right]\right),   
\end{equation}
where $I[A]$ is the Chern-Simons action written in a 2+1
decomposition,
\begin{equation}
I\left[A\right] = \frac{k}{4\pi} \int_M
\varepsilon^{ij}\Tr\left(-A_i\dot{A}_j + A_0 F_{ij}\right)d^3x.
\label{CS1}
\end{equation}
The coupling constant $k$ is given by
\begin{equation}
k=-\frac{l}{4G},
\label{k}
\end{equation}
just as in the Lorentzian case.

\subsection{The Euclidean 3d black hole and its complex structure}

The Euclidean black hole solution is obtained by defining $t=-it_E$
and $r_-=i\alpha$ in (\ref{mm19}), giving
\begin{equation}
ds^2 = \sinh^2\rho\left({r_+ dt_E\over l}-\alpha d\phi\right)^2 + 
l^2d\rho^2 +\cosh^2\rho\left({\alpha
dt_E\over l} + r_+ d\phi\right)^2.
\label{mm19eu}
\end{equation}
For the Euclidean calculation it is helpful to change coordinates to
\begin{equation}
  \label{max1}
  \varphi = \phi + \Omega t_E,\qquad x^0 = {t\over\hbar\beta},
\end{equation}
where
\begin{equation}
\beta = \frac{2\pi l^2 r_+}{\hbar(r_+^2 - r_-^2)}, \spc \Omega =
i\Omega_M =
-\frac{\alpha}{lr_+}.
\label{BO}
\end{equation}
Here, $\Omega_M=r_-/lr_+$ is the Minkowskian angular velocity.

The angular coordinate $\varphi$ has the standard period
$0\leq\varphi<2\pi$, while the time coordinate $x^0$, which is also
periodic, has the range $0\leq x^0<1$. The $\rho=const.$ surfaces in
the black hole manifold have thus the topology of a torus with the
identifications
\begin{equation}
\varphi\sim\varphi+2\pi n, \spc x^0 \sim x^0+m,
\label{id}
\end{equation}
with $n,m$ integers.  The radial coordinate $\rho$ has the range
$0<\rho<\infty$, with $\rho=0$ representing the black hole horizon.
Thus the Euclidean black hole manifold is represented by a solid
torus. The line $\rho=0$ represents the horizon, and is a circle at
the centre of the solid torus. We shall discuss below whether in the
sum over metrics in the partition function, this line should be
regarded as an inner boundary of the solid torus (see Fig. 1).

\begin{figure}
  \centerline{ \psfig{figure=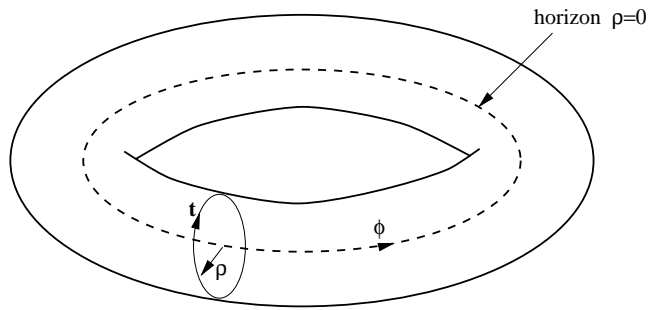,angle=-90,height=4cm}}
\caption{\sf The Euclidean black hole topology.}
\label{fig1a}
\end{figure}

The gauge field representing the Euclidean black hole is given by
\begin{eqnarray}
A^1 &=& \frac{(r_+ - i\alpha)\sinh\rho}{l} 
\left(d\varphi + \tau dx^0 \right),\nonumber\\
A^2 &=& id\rho, \label{Abh}\\
A^3 &=& \frac{i(r_+ - i\alpha)\cosh\rho}{l} 
\left(d\varphi + \tau dx^0 \right),\nonumber
\end{eqnarray}
where $\tau$ is a complex dimensionless number given by,
\begin{equation}
\tau = \hbar \beta \left( \Omega + \frac{i}{l}\right).
\label{tau}
\end{equation}
The corresponding formulae for $\bar A$ are obtained simply by complex
conjugation.

It is now natural to define a complex coordinate $z$ by
\begin{equation}
z = \varphi + \tau x^0.
\end{equation}
The identifications (\ref{id}) induce in the complex plane the
identifications
\begin{equation}
z \sim z + 2\pi n + \tau m , \spc n,m \mbox{~integers}.
\label{ics}
\end{equation}
We thus find that the black hole has a natural complex structure with
a modular parameter $\tau$ given by (\ref{tau}).

The important point is that by the coordinate transformation
(\ref{max1}), we have introduced a second pair of parameters into the
metric, which by virtue of the periodicity relations (\ref{id}),
should be thought of as the intensive parameters $\beta$ and $\Omega$.
On-shell, they are related to $r_\pm$ by (\ref{BO}), conditions which
emerge by either imposing the absence of conical singularities in the
Euclidean manifold or using the first law $\delta M = T \delta S +
\Omega \delta J$. Off-shell, $\tau$ and $r_\pm$ can be taken to be
independent.  

\subsection{Boundary conditions and boundary terms}

\subsubsection{Spatial infinity}

Let us first consider the outer boundary of the solid torus which
represents spatial infinity. The correct boundary term at infinity
should be consistent with boundary conditions that fix $\beta$ and
$\Omega$ at infinity.  In the complex coordinates $(z,\bar z)$, the
on-shell gauge field (\ref{Abh}) and its complex conjugate have the
property that
\begin{equation}
A_{\bar z}=0, \spc  \bar A_z =0.
\label{b11}
\end{equation}
In terms of the spacetime coordinates $x^0,\varphi$, these conditions
read,
\begin{equation}
A_0 = \tau A_\varphi, \spc  \bar A_0 = \bar\tau \bar A_\varphi,
\label{b1}
\end{equation}
as can be verified using $A_\varphi d\varphi + A_0 dx^0 = A_z dz +
A_{\bar z} d\bar z$ and that Im$(\tau) = \hbar\beta/l\neq 0$.
Eqs. (\ref{b11}), or (\ref{b1}), are just the Euclidean version
of the chirality conditions (\ref{chiral}) used in Sec. 2, and we
shall also use them here as part of our boundary conditions. The
residual gauge group is generated by the Kac-Moody currents ($\bar
A_{\bar z}$) $A_z$ which are (anti-) holomorphic functions of ($\bar
z$) $z$.

The main difference with the analysis of Sec. 2 is that we shall not
fix the value of $\int \mbox{Tr} A^2$ because we now work in a
grand-canonical ensemble. We thus need to add to the action
(\ref{CS1}) a boundary term at infinity in order to make it well
defined. The appropiate action for the grand-canonical ensamble
defined by the boundary conditions (\ref{b1}) and a fixed $\tau$ (note
that this fixes $\beta$ and $\Omega$ at infinity) is
\begin{equation}
I[A_i,A_0,\tau] = 
\frac{k}{4\pi} \int_M \epsilon^{ij} \mbox{Tr}\,
( -A_i \dot A_j + A_0 F_{ij} )- \frac{k\tau}{4\pi} 
\int_{T^2_\infty} \mbox{Tr}  A^2_\varphi \, ,
\label{CS-H}
\end{equation}
plus a boundary term at the horizon that we discuss in the next
paragraph.  [Note that conditions (\ref{b1})  were also used in
\cite{bg}, with a slight modification that makes the modular parameter
time dependent, in an attempt to obtain a better understanding of
Carlip's original paper \cite{Carlip1}.]

Using the expressions (\ref{m225}) for the mass and spin, it is easy
to see that the boundary term (the only term that survives in the
semiclassical limit) gives the correct weight factor
$-i\hbar\beta(E-\Omega J)$ for the grand-canonical ensemble.

\subsubsection{The horizon}

The boundary conditions at the horizon are more subtle. In the
Lorentzian theory it is quite natural to introduce a boundary term at
the horizon, since only the `outer' part of the black hole may be
viewed as physical. In the Hamiltonian formulation this means that one
has to fix the hypersurfaces at the bifurcation point which leads to
an isolated, non-smooth boundary, often referred to as joint or edge.
In \cite{Hayward} it was shown that such a non-smooth boundary gives
rise to additional boundary terms. In the case of black hole
spacetimes this joint contribution at the bifurcation point is
responsible for the appearance of a non-zero charge located at the
horizon, equal to one quarter the black hole area, which then can be
interpreted as the entropy of the black hole \cite{Brotz,Teitelboima}.  

In the Euclidean spacetime the situation is different since the
`inner' part of the black hole is already cut off from the manifold.
Nevertheless, it has been argued (see, for example,
\cite{Teitelboima}) that one also has to introduce a boundary at the
origin of the Euclidean spacetime by removing a point from the
Euclidean $(r,t)$-plane because the foliation using the vector field
$\partial_t$ is not well defined at $r=r_+$.  Since we are using a
Hamiltonian action, the $t=const$ surfaces are annuli with two
boundaries and it is necessary to give some boundary conditions at the
horizon in order to ensure that the Hamiltonian is a well defined
functional and its derivatives exist.

We must now decide which boundary conditions to impose at the horizon.
Note firstly that the on-shell black hole field (\ref{Abh}) has a time
component at $\varrho=0$ that does not vanish,
\begin{equation}
\label{timecoordinate}
A^3|_{\varrho=0} = -2\pi dx^0.
\end{equation}
Since $x^0$ is an angular coordinate, one may suspect that this
indicates the presence of a non-trivial holonomy in the temporal
direction. However, an explicit calculation using
(\ref{timecoordinate}) reveals that this is not the case and $A^3_0$
can be set equal to zero by a globally well-defined gauge
transformation.  The situation changes if one allows a conical
singularity at the horizon as advocated in \cite{ct}. In this case,
the non-vanishing of $A_0$ at the horizon does imply the existence of
a holonomy. To handle this situation classically, one can either
remove the horizon from the manifold and hence change the topology or,
alternatively, one can introduce a source term or Wilson line along
the horizon and work with a solid torus topology. The introduction of
a Wilson line at the core of the solid torus was also suggested in
\cite{ct}, but as far as we know its consequences were never explored.

The above discussion suggests that we should look to fix $A_0$ at the
horizon. This can be achieved using the canonical action with no
boundary term. However, this does not then fix the condition
(\ref{timecoordinate}) (as opposed to allowing conical singularities).
This can be ensured only by introducing a Wilson line term whose
variation, when coupled to the bulk action, fixes
(\ref{timecoordinate}). The Wilson line term is 
\begin{equation}
\frac{k}{4\pi} \int_{S^1} \Tr \left(K \dot{X^\mu}{A_\mu} - 
    \bar{K}\dot{X^\mu}{\bar{A}_\mu}\right)
  \delta^3(X^\mu(\tau)-x^\mu)d\tau dx^0 d^2x,
\label{z2}
\end{equation}
which is localised along a non-contractible loop in the solid torus
defined by $X^\mu(\tau)$. Here $\tau$ is a parameter along the Wilson
line. $K$ is an element in the Lie algebra of the group that specifies
the vector charge of the source. We could also have included a
dynamical source with its own kinetic term
\cite{bmss,carlip,Witten89,ok}, but it is not clear that this is
necessary.

To conform with the usual choice of coordinates, we shall take the
Wilson line to be located along the curve $\rho=0$, and to be
parametrised by the variable $\varphi$. Since we want the action
principle to contain the black hole in its space of solutions, the
strength of the source is fixed by looking at the classical field
(\ref{Abh}) at the point $\rho=0$.  We take
\begin{equation}
  \label{m1}
  K^a=\left(0,0,-4\pi\right),
\end{equation}
and the source term becomes
\begin{equation}
  \label{m2}
  {k/2} \int_M \left(A^3_\varphi - 
    \bar{A}^3_\varphi\right)\delta^2(\underline{x})d\varphi d^2x,
\end{equation}
where now $d^2x$ and $\delta^2(\underline{x})$ refer to the $\rho,t$
plane (not $\rho,\varphi$). Note that the choice of orientation of $K$
in the Lie algebra is unimportant, since it only fixes the orientation
of the solution in internal space. The source term (\ref{m2}) is the
analogue of the geometrical horizon term proportional to the area that
was mentioned above. Indeed, on-shell, the value of this term (plus
the other copy) is equal to one fourth of the area.  In the off-shell
language of Carlip and Teitelboim \cite{ct}, the field equations will
lead to $\Theta=2\pi$ and $\Sigma=0$ as expected, since no conical
singularities are allowed.

\subsubsection{The action}

The total Euclidean action for the black hole is therefore the sum of
(\ref{CS-H}) and (\ref{m2}) giving,
\begin{equation}
I[A,\tau; \bar A,\bar\tau] = I[A,\tau] - I[\bar A,\bar\tau] ,
\end{equation}
with
\begin{eqnarray}
  \label{y1}
I[A,\tau] &=& 
\frac{k}{4\pi} \int_M \epsilon^{ij} \mbox{Tr}\,
\left( -A_i \dot A_j + A_0 F_{ij} \right)
\cr
&&
- \frac{k\tau}{4\pi} 
\int_{T^2_{\rho=\infty}} \mbox{Tr} A^2_\varphi 
 + \frac{k}{2} \int_{S_{\rho=0}^1} A^3_\varphi ,
\end{eqnarray}
(where since $\varepsilon^{tij}=\varepsilon^{ij}$, in Euclidean space,
the sign of the bulk term is different to that in (\ref{a1})). It is
worth stressing here, once again, that this action has well defined
variations provide $\tau$ is fixed.

Note, finally, that the micro-canonical action (\ref{a1}) and the
grand-canonical action (\ref{y1}) differ by exactly the boundary term
at infinity equal to $\beta(E+\Omega J)$ that one would expect on
general grounds, and that has been advocated by Brown and York
\cite{by}.

We shall now explore the semi-classical and quantum mechanical
consequences of this action.

\subsection{Semiclassical partition function}

The action (\ref{CS-H}) with the source term (\ref{m2}) has the right
semiclassical value. Since the canonical action is zero on the
classical black hole background (\ref{Abh}) one obtains,
\begin{equation}
Z_{\mbox{\small semiclassical}} = e^{-\beta (M + \Omega J) + S},
\label{Zsc}
\end{equation}
where $S$ is given by
\begin{equation}
S = \frac{2\pi r_+}{4 \hbar G},
\end{equation}
as expected, and comes entirely from the source term at the horizon.

This partition function is grand-canonical because in the action
principle only $\beta$ and $\Omega$ (or $\tau$) were fixed. This means
that $Z$ is a function of $\beta$ and $\Omega$. Using (\ref{mm19a})
and (\ref{BO}) one can write $Z$ as a function of $\beta$ and
$\Omega$,
\begin{equation}
Z(\beta,\Omega) = \exp\left[\frac{\pi^2 l^2}{2 \hbar^2 G \beta
(1+l^2\Omega^2 )}\right].
\label{GH}
\end{equation}
This is the semiclassical value of the grand-canonical partition
function.

\subsection{The partition function and the chiral WZW model}

In a Chern-Simons formulation of three dimensional Euclidean quantum
gravity, the partition function involves a sum over an $SL(2,\C)$
gauge field.  We must therefore deal with the fact that the group
$SL(2,\C)$ is not compact and the black hole manifold has a boundary.
(For manifolds without boundaries it has been proved in
\cite{Witten91} that $Z$ can be understood as a complexified $SU(2)$
problem.) As has been stressed in \cite{Carlip2}, one can hope to make
progress by treating each of the complex connections $A$ and $\bar{A}$
as real, so that the partition function becomes just the product of
two, complex conjugate, $SU(2)$ partition functions. Following
\cite{Carlip2} we shall write
\begin{equation}
Z = |Z_{SU(2)}|^2,
\label{z1}
\end{equation}
and compute $Z_{SU(2)}$, hoping to make sense of  its relation to
the trace over states in $SL(2,C)$ by some
form of analytic continuation.

We thus consider the functional integral,
\begin{equation}
Z_{SU(2)}(\tau) = \int D[A_i]D[A_0] 
\exp\left(\frac{i}{\hbar} I[A_i,A_0;\tau]\right),
\end{equation}
where $I$ is given in (\ref{y1}), and we integrate over all gauge
fields satisfying the boundary conditions (\ref{b1}).

Integrating over $A_0$ gives the constraint that $F_{ij}=0$, which
implies that $A_i = h^{-1} \pp_i h$ where $h$ is a map from $M$ to the
group. We also have to fix a gauge and we can do this using the gauge
fixing choice (\ref{a8a}) which fixes
\begin{equation}
  \label{zz1} h(\rho,\varphi,t) = b(\rho) g(\varphi,t),\qquad
  A(\varphi,t) = g^{-1} \pp_\varphi g.
\end{equation}
Since the first homotopy group of the solid torus is non-trivial, $g$
could be multi-valued.  After inserting the gauge fixed and flat $A_i$
into the functional integral one obtains an expression that only
depends on the boundary values of the map $g$ \cite{Moore-Seiberg},
\begin{equation}
Z_{SU(2)}(\tau) = \int Dg \exp\left(
\frac{i}{\hbar} I_{CWZW}[g,\tau]\right), 
\end{equation}
where the chiral WZW action is given by,
\begin{eqnarray}
I_{CWZW}[g,\tau] &=& -\frac{k}{4\pi} 
\int_{T^2} \mbox{Tr} (\pp_\varphi g^{-1} \dot g) - 
\frac{k}{12\pi} \int_M \mbox{Tr}(g^{-1}dg)^3  -  \nonumber\\ 
&& \frac{\tau k}{4\pi}\int \mbox{Tr} (g^{-1} \pp_\varphi g)^2 +
\frac{k}{4\pi} 
\int \Tr (Kg^{-1}\pp_\varphi g) .
\label{CWZW}
\end{eqnarray}
Here $K$ is related to the original $K$ of (\ref{m1}) by conjugation
by $b(\rho)$ and so may be taken equal to $K$ without loss of
generality.

The reduction of the three dimensional problem to a two dimensional
conformal theory is a consequence of the absence of propagating
degrees of freedom in the three dimensional field theory. This allowed
us to solve the constraint.  A second consequence of the absence of
degrees of freedom is that the boundary term at the horizon is now
linked to the boundary term at infinity. The conformal field theory
lives on a torus with no reference at all to the radial coordinate.

The chiral WZW action (\ref{CWZW}) has two pieces. The kinetic term
(first line) defines the commutation relations of the theory. As is
well known \cite{Witten84}, these commutation relations are given by
the $SU(2)$ Kac-Moody algebra,
\begin{equation}
~[T^a_n,T^b_m ] = i \hbar \epsilon^{ab}_{\ \ c} T^c_{n+m} 
                   + n \hbar \frac{k}{2} 
\delta^{ab} \delta_{n+m,0},
\label{km}
\end{equation}
where the $T^a_n$ are the Fourier components of the gauge field,
\begin{equation}
A_\varphi = g^{-1} \pp_\varphi g = \frac{2}{k} 
\sum_{n=-\infty}^\infty T^a_n e^{in\varphi}.
\end{equation}
Note that it is more convenient to define the $T^a_n$ in this way
rather than as in (\ref{a10}) in Euclidean space.  The second piece
(second line in (\ref{CWZW})) is the Hamiltonian.  Since the
Hamiltonian involves $A^2$ it has to be regularised by choosing a
normal ordering. Moreover, it is well known that the coefficient of
$L_0$ in the non-Abelian theory is not $k^{-1}$ but rather
$(k+\hbar)^{-1}$. In the following we shall be interested in the limit
of large $k$ and therefore this shift can be neglected.

The partition function can then be calculated as
\begin{equation}
Z(\tau) = \sum_{2s=0}^{k/\hbar} \mbox{Tr}_s\
\exp\left(\frac{i}{\hbar}\tau
L_0 
+ 2\pi \frac{i}{\hbar} T_0^3 \right),
\label{Z0}
\end{equation}
where for large $k$, $L_0$ is given by
\begin{equation}
L_0 = \frac{1}{k} \sum_{n=-\infty}^{\infty} :T^a_{-n} T^b_{n} :
\delta_{ab},
\end{equation}
and $s$ labels the spin of the different $SU(2)$ representations (it
can be integer or half-integer). The symbol Tr$_s$ represents a trace
over states belonging to the representation with spin $s$.  The spin
structure term can be thought of as being equivalent to having a
non-zero flux through the hole created by closing the time direction.

The problem of computing $\mbox{Tr}_s\,\left( q^{L_0/\hbar} e^{i\theta
    T_0^3/\hbar} \right)$ for a given value of $s,q$ and $\theta$ is
well known and explicit formulae are available.  For $SU(2)$, writing
$q=e^{i\tau}$ and taking $\theta=2\pi$, one has
\cite{Goddard-Kent-Olive},
\begin{equation}
\mbox{Tr}_s \left(q^{L_0/\hbar}e^{i2\pi T_0^3/\hbar}\right) =
  \frac{ q^{\hbar s(s+1)/k}  \sum_{n=-\infty}^\infty 
(-1)^{2s+2 n k/\hbar} (2s+1 + 2 n k/\hbar)
    q^{{n^2 k/\hbar }+ (2s+1)n}}{\Pi_{m=1}^\infty (1-q^m)^3} .
\label{ch}
\end{equation}

The denominator in (\ref{ch}) does not depend on $s$ and it is
therefore a global factor in the partition function. This factor
provides a quantum correction to the entropy that does not depend on
Newton's constant.  The value of this contribution can be calculated
in the limit of small $\tau$ (large black holes) as
\begin{equation}
\Pi_{m=1}^\infty (1- q^m) \approx e^{\pi^2/6(q-1)} .
\end{equation}
Inserting $q=e^{i\tau}$ and using (\ref{S}) and (\ref{z1}), the
contribution to the entropy from this term is equal to
\begin{equation}
S_0 = \frac{\pi r_+}{l}. 
\end{equation}
This correction, which does not depend on $G$, has already appeared in
the literature \cite{ct,Carlip2,Lee}.

Let us now consider the numerator in (\ref{ch}). In the limit of large
$k$, the sum over $n$ is suppressed because $e^{-k n^2 \beta/l}
\rightarrow 0$ exponentially for $n \neq 0$.  We thus keep only the
term $n=0$. This means that the Bekenstein--Hawking entropy does not
come from the higher Kac-Moody modes but from the sum over
representations. This is quite different from the analysis in
\cite{Carlip2} in which the entropy comes from the term with zero
spin\footnote{Note, however, that it is possible that the source term
  at the horizon could be regarded as an effective action term that
  arises from integrating out modes that are not seen in this
  $SU(2)\times SU(2)$ calculation.}.  Setting $n=0$ and defining
$j=2s$, we arrive at
\begin{equation}
Z_{SU(2)} = Z_0 Z_{1/k} \sum_{j=1}^{k/\hbar} j 
\, (-1)^j \, e^{i\hbar\tau  j^2/4k} ,
\label{S1}
\end{equation}
where $Z_0$ represents the correction that does not depend on $G$,
while $Z_{1/k}$ the contribution from the sum over $n$ whose logarithm
vanishes at least as $1/k$.

Note that in the quantum calculation the term $\exp\left(2\pi i
  T_0^3\right/\hbar)$, arising from the boundary term at the horizon,
produces the factor $(-1)^{2s}$ which may be interpreted as a $(-1)^F$
operator that alternates bosonic and fermionic representations. We
shall see that this operator has an important role in producing the
right contribution to the entropy.

The sum (\ref{S1}) is not what we want because the black hole does not
belong to the set of states that we are considering when we calculate
the $SU(2)$ partition function. Indeed, $j$ labels unitary $SU(2)$
representations for which
\begin{equation}
  \label{n1}
  L_0\ket{j} = {1\over k}T^a_0T^b_0\delta_{ab}\ket{j} = 
  {\hbar^2 j(j+2)\over
    4k}\ket{j}.
\end{equation}
However, the value of $L_0$ on the black hole
background is $L_0= - k(r_++i\alpha)^2 /4l^2$, from which it follows
that 
\begin{equation}
j^2 = -\frac{k^2 (r_++i\alpha)^2}{\hbar^2 l^2},
\label{j}
\end{equation}
for large $j$. Thus, the states we are interested in for the black
hole belong to an $SU(2)$ representation with complex spin. Even in
the non-rotating case, $\alpha=0$, $j^2$ is negative and thus cannot
be real. We shall not attempt to give any interpretation to such a
representation here, but the reason behind it lies in the fact that
Euclidean gravity is a Chern-Simons theory for the group $SL(2,\C)$
rather than two copies of $SU(2)$.

Let us analytically continue $j$ to the complex plane and set
$j \rightarrow i{\jmath}$.  The sum reduces to
\begin{equation}
Z_{SU(2)} = Z_0 Z_{1/k} \sum_{\jmath} {\jmath}\, \exp\left(-{i\hbar
\tau 
    \jmath^2\over 4 k} + \pi {\jmath}\right).  
\label{S2}
\end{equation}
Note that the $(-1)^F$ operator now has eigenvalues $e^{\pi
{\jmath}}$.
Remarkably, this term which produced the right entropy in the
semiclassical calculation, also provides the right degeneracy in this
quantum mechanical calculation. 

Consider the total partition function $Z=|Z_{SU(2)}|^2$. Since
$\jmath$ is complex we define $\jmath=j_1+ij_2$. Using $\tau =
\hbar\beta(\Omega+i /l)$ and $k=-l/4 G$ we find
\begin{equation}
  Z = |Z_0 Z_{1/k}|^2 \sum_{{j_1}, {j_2}}  \left(j_1^2+j_2^2\right)\,
  \exp\left( -\beta (M_{j_1,j_2} + \Omega J_{j_1,j_2})  
  + 2\pi {j_1} \right).  
 \label{S4}  
\end{equation}
with
\begin{equation}
M_{j_1,j_2} = \frac{2G\hbar^2  (j_1^2-j_2^2)}{l^2}, \ \ \ \ \ \ \
J_{j_1,j_2} = \frac{4G\hbar^2  j_1 j_2}{l}
\label{MJj}
\end{equation}
Since, for a black hole, $M$ and $J$ are related to the inner and
outer horizons by (\ref{mm19a}) (or in the Euclidean version by
replacing $r_-=i\alpha$), we obtain
\begin{equation}
r_+ = 4G\hbar j_1, \ \ \ \ \ \ 
\alpha = 4G\hbar j_2. 
\label{r1}
\end{equation}

The term $(j_1^2+j_2^2)$ outside the exponential in (\ref{S4})
combines with the sum to give the required measure over $M$ and $J$.
The entire partition function becomes
\begin{equation}
Z(\beta,\Omega) \sim {l^3\over 16\hbar^4 G^2}
\int dM dJ \, \rho(M,J)\, \exp\left(-\beta(M+\Omega J)\right)
\end{equation}
and implies that the density of states is $\rho(M,J) = \exp (2\pi
j_1)$. Using (\ref{r1}), we obtain
\begin{equation}
  \label{n5}
  \rho(M,J) = \exp\left(2\pi r_+(M,J) \over 4\hbar G\right),
\end{equation}
in complete agreement with the
Bekenstein--Hawking value. 

Finally, the semiclassical grand-canonical partition function
(\ref{GH}) and thus the entropy $S$ can be obtained by a simple saddle
point approximation (\ref{S2}).  Noticing that the sum (\ref{S2}) has
a saddle point at ${\jmath}= 2\pi i k /\hbar \tau$ we find that
\begin{equation}
Z_{SU(2)}(\tau) = Z_0\, Z_{1/k}\,\exp\left(
\frac{i\pi^2 l}{4\hbar G\tau}\right),
\end{equation}
where we have inserted $k=-l/4 G$. Computing the complex
modulus of $Z$, and taking into account the value of $\tau$ given in
(\ref{tau}), we find that
\begin{equation}
Z(\beta,\Omega) = |Z_0\, Z_{1/k}|^2 \exp\left(\frac{\pi^2 l^2 }{2
\hbar^2G 
\beta (1+l^2\Omega^2)} \right),
\end{equation}
in complete agreement with (\ref{GH}).

It is interesting to note that this calculation can be repeated in the
case where $k\to\infty$, the semiclassical limit, in a purely abelian
theory. Details of this calculation are given in the appendix.

\section{Conclusions}
\setcounter{equation}{0}

We have performed two separate calculations of the entropy of the 2+1
dimensional black hole using the relation between 2+1 dimensional
gravity and Chern-Simons theory. In Sec. 2, we have worked in the
micro-canonical ensemble, and have calculated the density of states
starting from the Kac-Moody algebra of global charges (WZW theory).
We have computed the correct density of states by relating the global
charges to a  particular Virasoro algebra via a twisted Sugawara
construction, in a way first considered in \cite{max}. This Virasoro
algebra turns out to generate the same asymptotic isometries
considered in Refs.  \cite{bh,Strominger}, if the analysis of global
charges is performed at infinity. We have shown that it is also
present on any other boundary surface at constant radius, including
the black hole horizon.  In Sec. 3, we have worked in the
grand-canonical ensemble, which we have defined by adding an
appropriate boundary term at infinity. We have found that in order to
obtain the correct partition function we must also add a source term
at the horizon. This source term gives the correct value of the
partition function both semi-classically and in an exact quantum
mechanical calculation. It is the analogue of the term equal to $A/4G$
that is sometimes added to the canonical Einstein--Hilbert action to
yield the correct semi-classical partition function for black holes in
arbitrary dimensions.

In the micro-canonical calculation we saw that we obtain the correct
density of states at a given value of mass and spin after we make a
reduction from the WZW theory to a theory of boundary deformations
that satisfies the Virasoro algebra, or equivalently to a Liouville
theory.  Since this reduction involves additional constraints on
  the allowed global charges, it one expects that the density of
  states should be greater in the WZW theory (although possibly equal
  to leading order). Why, then, can one not calculate the density of
  states directly in terms of representations of the Kac-Moody
  algebra?  In the WZW theory, it seems clear that there are an
  insufficient number of states (and for this reason, in the
  grand-canonical ensemble the correct partition function required the
  addition of a source term to give a larger apparent degeneracy). The
  answer presumably lies in the use of a twisted Sugawara construction
  to connect the Kac-Moody and Virasoro algebras.  Although the states
  we eventually count are unitary states with respect to the standard
  quantisation of the Virasoro algebra, they most probably correspond
  to a non-unitary, twisted representation of the Kac-Moody
  algebra\footnote{Carlip \cite{carlipwhat} has pointed out that
      the correct number of states is obtained through a
      representation of the Virasoro algebra with $c=3l/2G$ only if
      the vacuum has $L_0=0$, which translates to a negative
      eigenvalue for $\tilde{L}_0$ defined in an untwisted Sugawara
      construction from the underlying WZW theory. However, it is
      interesting to note from (\ref{m224}) that the condition $L_0=0$
      corresponds to $M=-1/8G$, $J=0$ (Anti-de Sitter space) and
      $\tilde{L}_0=-l/16G$, while the black hole vacuum $M=J=0$ has
      $\tilde{L}_0=0$ but $L_0=l/16G$. This strongly suggests that our
      ensemble includes Anti-de Sitter space as its true vacuum.}.
  Thus in order to find the correct density of states, we should look
  at a different set of states to those usually constructed in
  representations of the Kac-Moody algebra. Of course, this
  calculation is further complicated by the fact that representations
  of $SL(2,R)$ WZW theory are poorly understood.

In contrast to the micro-canonical case, in the grand-canonical
calculation we have managed to obtain the correct density of states
directly in a standard (not twisted) WZW theory, using standard
expressions for the partition function (and an analytic continuation).
However, this result came from a partition function with a
``spin-structure'' term, twisting the WZW theory in the time
direction. It seems likely
that the twists in the space and time directions that we have
discussed are related by a modular transformation\footnote{It would be
interesting to compute the Euclidean partition function in a
canonical framework where $\varphi$ is the time coordinate and to
verify that this can yield the same partition function.}.

In this context, it is also interesting to speculate on the correct
interpretation of the source or Wilson line term that gives rise to
the non-trivial ``spin-structure''. We saw in our semiclassical and
quantum mechanical calculations that this term produces the black hole
entropy not as a density of states, but rather as an operator
eigenvalue. We conjecture that this source term can be understood in a
different context as an effective action term. If we begin with the
micro-canonical picture of a twisted WZW model with trivial
spin-structure, then it should be possible to get back to an untwisted
WZW theory by integrating out the additional degrees of freedom
arising from the spatial twisting of the WZW theory. The effect of
integrating out these states would be to introduce the spin-structure
term. It would be extremely nice to see this connection
explicitly. 

Finally, we comment on what these various calculations tell us about
the location of the degrees of freedom giving rise to the black hole
entropy. While Carlip \cite{Carlip1,Carlip2} has advocated that these
degrees of freedom should be located on the black hole horizon,
Strominger \cite{Strominger} has shown that the algebra of asymptotic
isometries of the metric leads to the correct density of states. Our
discussion of the global charges and of the reduction to Liouville
theory in Sec. 2 showed explicitly that the Virasoro algebra
responsible for the density of states can live at any value of the
radius $\rho$. A similar conclusion is suggested by the $\rho$
independence of the grand-canonical calculation. We were able to
obtain an explicit form for a set of deformations of the horizon whose
classical algebra is the Virasoro algebra, with the same classical
central charge as the set of asymptotic isometries, that gives the
correct Bekenstein--Hawking entropy. 

The discovery of an algebra of operators at the horizon, whose
representations yield the correct density of states, makes an
extension of these ideas to higher dimensions look more plausible.
Whereas it is unlikely that the algebra of asymptotic isometries of
black holes in higher dimensions could lead to the correct density of
states, it seems more likely that an algebra of deformations of the
horizon could have the required properties.  Although these two
algebras are identical in our case, this is because of the trivial
dynamics of 2+1 dimensional gravity and would certainly not be true in
general. In higher dimensional applications, one would then have to
relate these charges at the horizon to the mass and spin of the black
hole. This problem is solved in our case by the same trivial dynamics
of the theory.

What is not so clear from our analysis is the physical interpretation
of the subset of deformations that lead to the Virasoro algebra at any
finite radius. However, as mentioned above in the context of WZW
theory, the restriction to this subset of deformations will reduce the
density of states, but it may well not change it to the leading
semi-classical order. In that case, the complete algebra of
deformations of the horizon would lead to the correct density of
states. However, checking this probably requires getting a handle on
the problem of state counting in the WZW theory.

We are hopeful that a generalisation to arbitrary dimensions of the
calculations that have been developed for the 2+1 dimensional black
hole may come about through the algebra of deformations of the
horizon. Indeed, this algebra may have a very direct application for
black holes in higher dimensions whose near-horizon behaviour is
similar to the 2+1 dimensional black hole. In this case we would be
able to talk of states localised at the horizon.

\section*{Acknowledgements}
We thank R. Baeza, S. Carlip, F. Falceto, A.Gomberoff, M. Gaberdiel,
M. Henneaux, T. Jacobson, C. Mart\'{\i}nez, F. Mendez, I. Sachs, S.
Stanciu, C.  Teitelboim, A. Tseytlin, L. Vergara and J. Zanelli for
helpful discussions.  MB was supported by grants \#1970150 from
FONDECYT (Chile), CICYT (Spain) \#AEN-97-1680, and also thanks the
Spanish postdoctal program of Ministerio de Educaci\'on y Cultura.  TB
acknowledges
financial support from the German Academic Exchange Service (DAAD).
MEO was supported by the PPARC, UK.

\appendix

\section{The Abelian WZW Theory}
\setcounter{equation}{0}

In the weak coupling limit the non Abelian nature of the theory can be
neglected. Therefore it should also be possible to derive the
semiclassical value of the partition function from the Abelian
theory, and we shall now show that this is indeed the
case. If we define
\begin{equation}
\label{c0} g(t,\varphi) = e^{X^a(t,\varphi) J_a} ,
\end{equation}
where $J_a$ are now the generators of the Abelian Lie algebra, the
chiral WZW action (\ref{CWZW}) reduces to
\begin{equation}
\label{c1}
I_{CWZW}[X,\tau]=\frac{k}{8\pi} \int \left( -\partial_0 X^a
  \partial_\varphi X_a + \tau \left(\partial_\varphi X\right)^2 - 4\pi
  \partial_\varphi X^3 \right).   
\end{equation}
{From} this action we find the field equations
\begin{equation}
\label{c2} (\partial_0 - \tau \partial_\varphi) \partial_\varphi X^a
= 0.   
\end{equation}
Thus the general solution of these field equations contain only right
moving modes which expresses the chirality of our theory.  Motivated
by the field equation we may expand $X^a$ in normal modes as,
\begin{equation}
\label{c11} k X^a(x^0,\varphi) = 2 \varphi \alpha_0^a(x^0)+ 2 \sum_{n
\neq
  0} \frac{\alpha_n^a (x^0)}{in} e^{in\varphi} + 2 \alpha(x^0).   
\end{equation}

Note that the action (\ref{c1}) has a gauge symmetry, $\delta X^a =
\epsilon(x^0)$, which we use to set the function $\alpha(x^0)$ to
zero.  Replacing the mode expansion into the chiral WZW action, one
obtains
\begin{equation}
\label{c13} 
I_{CWZW}[\alpha]= \int dx^0 \left( - \sum_{n\neq 0}
  \frac{ \dot{\alpha}_0^a \alpha_{a\:n}}{ikn} + \sum_{n\ge 1} \frac{2
    \dot{\alpha}_{-n}^a \alpha_{a\:n}}{ikn}+ \tau L_0 - 2\pi
  \alpha_0^3 \right),  
\end{equation}
where $L_0$ is a Virasoro generator,
\begin{equation}
\label{c14} 
L_0= \frac{1}{k} ( \alpha_0^2+2 \sum_{n\ge 1} \alpha^a_{-n}
\alpha_{an} ),  
\end{equation}
and where we have eliminated total derivative terms. The action
(\ref{c13}) gives rise to the Abelian Kac-Moody algebra
\begin{equation}
\label{kaka} 
\left[ \alpha_n^a, \alpha_m^b \right] = n \hbar
\frac{k}{2} \delta^{ab} \delta_{n+m,0} , 
\end{equation}
from which we can define a Fock space in the standard manner.

The partition function can be calculated as,
\begin{equation}
\label{cz}
Z(\tau) = \int D[\alpha] \exp \left( \frac{i}{\hbar}
I_{CWZW}[\alpha,\tau]
\right) 
 = \int d\alpha_0 \Tr \exp\left( \frac{i\tau}{\hbar} L_0
 -\frac{i}{\hbar}
2\pi \alpha_0^3 \right). 
\end{equation}
{From} the field equation (\ref{c2}) we deduce again that
semi-classically only the zero modes contribute to the partition
function. Hence we may split the partition function (\ref{cz}) into a
leading and a sub-leading part
\begin{equation}
\label{c16} 
Z(\tau) = \Tr q^N \int
d\alpha_0 \exp\left\{\frac{i}{\hbar} \left( \tau \frac{\alpha_0^a
      \alpha_{0\: a}}{k} - 2\pi \alpha_0^3\right) \right\},  
\end{equation}
where we have introduced the number operator
\begin{equation}
\label{c16a} 
N= \frac{1}{\hbar k}
\sum_{n=1}^{\infty} \delta_{ab} \: \alpha^a_{-n} \alpha^b_n.  
\end{equation}
In the saddle point approximation the integral over $\alpha_0$
($\alpha_0^1 = \alpha_0^2=0$ and $\alpha_0^3 = k \pi/ \tau$) is easily
evaluated
\begin{equation}
\label{c18}
Z(\tau) = \Tr q^N \exp \left\{ \frac{\pi^2 l^2}{4 \hbar^2 G\beta
    (1-il\Omega)}\right\}   .
\end{equation}
Note that the integral over $\alpha_0$ is only well defined, if we
assume $\alpha_0^2<0$. For the calculation of the prefactor we need to
know the number of states at each level $N=n$. This calculation is
well known and can be found in \cite{GrSchwWit}. In the limit $\tau
\rightarrow 0$ (large black holes) one finds
\begin{equation}
\label{c19} 
\Tr
q^N = \exp\left(\frac{i\pi^2}{2\tau}\right).   
\end{equation}
The total partition function is thus given by
\begin{equation}
\label{c20} 
Z(\beta,\Omega) = |Z(\tau)|^2 =
\exp\left(\frac{\pi^2 l^2}{2\hbar^2 G\beta(1+l^2\Omega^2)} +
\frac{\pi^2
    l}{\hbar \beta(1+l^2 \Omega^2)} \right).   
\end{equation}
Thus the Abelian calculation not only produces the Bekenstein- Hawking
part of the partition function but also leads to the $Z_0$ correction
discussed above.

\end{document}